\documentclass[12pt]{article}
\usepackage[utf8]{inputenc}
\usepackage[a4paper]{geometry}
\usepackage{url}
\usepackage{float}
\usepackage{appendix}
\usepackage[tbtags]{amsmath}
\usepackage{amssymb,amsfonts,bm,latexsym}
\usepackage{amsthm}
\usepackage{amsmath}
\usepackage{authblk}
\usepackage{bbm}
\usepackage{xcolor}
\usepackage{multirow}
\usepackage{listings}

\usepackage{afterpage}

\usepackage{graphicx}
\usepackage{subcaption}

\usepackage[authoryear]{natbib}
\usepackage{graphicx}

\title{On the performativity of SDG classifications in large bibliometric databases}
\author[1]{Matteo Ottaviani\footnote{Corresponding author, email: \texttt{ottaviani@dzhw.eu}}}
\author[1,2]{Stephan Stahlschmidt}

\affil[1]{\small{German Centre for Higher Education Research and Science Studies (DZHW), Schützenstr. 6a, 10117 Berlin (Germany)}}

\affil[2]{Unit of Computational Humanities and Social Sciences (U-CHASS), EC3 Research Group, University of Granada, Granada, Spain}
\date{\today}

\begin{document}

\maketitle

\begin{abstract}
Large bibliometric databases, such as Web of Science, Scopus, and \\OpenAlex, facilitate bibliometric analyses, but are performative, affecting the visibility of scientific outputs and the impact measurement of participating entities. Recently, these databases have taken up the UN's Sustainable Development Goals (SDGs) in their respective classifications, which have been criticised for their diverging nature. This work proposes using the feature of large language models (LLMs) to learn about the ``data bias" injected by diverse SDG classifications into bibliometric data by exploring five SDGs. We build a LLM that is fine-tuned in parallel by the diverse SDG classifications inscribed into the databases’ SDG classifications. Our results show high sensitivity in model architecture, classified publications, fine-tuning process, and natural language generation. The wide arbitrariness at different levels raises concerns about using LLM in research practice.
\end{abstract}

\medskip

\noindent \emph{Keywords}: Sustainable Development Goals, Classifications, Bibliometric databases, Large Language Models, Text Analysis, Noun Phrases, OpenAlex, Web of Science, Scopus.

\newpage

\section{Introduction}

Bibliometric databases play a critical role as digital infrastructures that enable bibliometric analyses and impact assessments within the scientific community. However, it is essential to acknowledge that these databases are not impartial; instead, they have a performative nature, as they are constructed based on specific understandings of the science system and value attributions \citep{whitley2000intellectual,vinkler1988bibliometric}.
Recently, there has been significant attention given to the contribution of the science system and its entities to the United Nations' Sustainable Development Goals (SDGs) in the bibliometric impact debate \citep{mishra2023bibliometric,meschede2020sustainable}. The SDGs provide a comprehensive framework for addressing global challenges and promoting sustainable development in various domains. The latter is a global framework that includes monitoring mechanisms and indicators. European countries have adapted their national sustainability indicator systems to align with the UN Agenda 2030. 
In the context of bibliometrics, SDG classifications are promoted to assess the societal relevance and impact of scientific research \citep{armitage2020mapping}. 

Major bibliometric databases, including Web of Science, Scopus, and OpenAlex, have introduced bibliometric classifications aligning publications with specific SDGs to facilitate the measurement of scientific contributions towards the SDGs. \cite{armitage2020mapping} carry out a bibliometric study aimed at comparing the Bergen and Elsevier approaches to finding scholarly publications related to the United Nations' SDGs. They show that the overlap in publications retrieved by the latter two approaches is small. Different search terms, combinations, and query structures significantly impact the retrieved publications, affecting the country rankings. 
The latter inconsistencies are not due to technical issues but rather different interpretations of SDGs. Understanding the reasons behind these differences is crucial for comprehending the performative aspects of bibliometric classifications and their impact on scientific outputs. We propose the application of Large Language Models (LLMs) for this purpose.

LLMs are pre-trained models utilizing deep learning techniques to generate human-like responses based on given prompts \citep{radford2019language}. These models are trained on vast amounts of text data and have shown remarkable language generation capabilities. Nevertheless, concerns have been raised about the objectivity and potential biases embedded within the generated answers \citep{bender2018data,lipton2018mythos}. 

In this work, we propose using LLMs, specifically the DistilGPT-2 model, a variant of GPT-2 particularly fitting our purposes \citep{radford2019language,sanh2019distilbert}, to gain insights into the qualitative biases introduced by diverse SDG classifications. 
The choice of this LLM is due to its great compromise between embedding no prior knowledge about SDGs ( DistilGPT-2 has been trained on a small dataset and, then, incorporated a significantly lower structural data bias compared to other renown LLMs) and serving basic LLM functions.
We examine the following five SDGs: SDG 4. Ensure inclusive and equitable quality education and promote lifelong learning opportunities for all; SDG 5. Achieve gender equality and empower all women and girls; SDG 8. Promote sustained, inclusive and sustainable economic growth, full and productive employment and decent work for all; SDG 9. Build resilient infrastructure, promote inclusive and sustainable industrialization and foster innovation; SDG 10. Reduce inequality within and among countries.

Our research design involves three main steps: data collection and analysis, fine-tuning the LLM for each bibliometric database and each SDG, and employing text analysis techniques to explore the biases and variations in the generated responses. We identify a \textit{jointly indexed publication} dataset of Web of Science, OpenAlex, and Scopus, counting 15,471,336 publications from 2015 to July 2023, which serves as the common ground for this research to identify the impact of the SDG classifications irrespective of the varying coverage of the underlying databases. From this dataset, we collect the varying publications attributed to the 5 SDGs mentioned above by each databases’ SDG classification, creating distinct publication subsets from the common ground set for the fine-tuning process. For each SDG, three fine-tuned LLMs are then administered the same collection of prompts, allowing us to compare and analyse the generated responses. 
That is we condense the publication-level differences inflicted by the diverse SDG classifications among millions of publications into an aggregate textual summary to uncover structural differences between the SDG classifications. The responses generated by the LLMs allow us to observe general patterns on the classification's substance that are otherwise not observable due to the sheer amount of classified publications. The text analysis reveals distinct linguistic features and varying perspective associated with each SDG classification.

\begin{figure}
\centering
\includegraphics[width=0.50\linewidth]{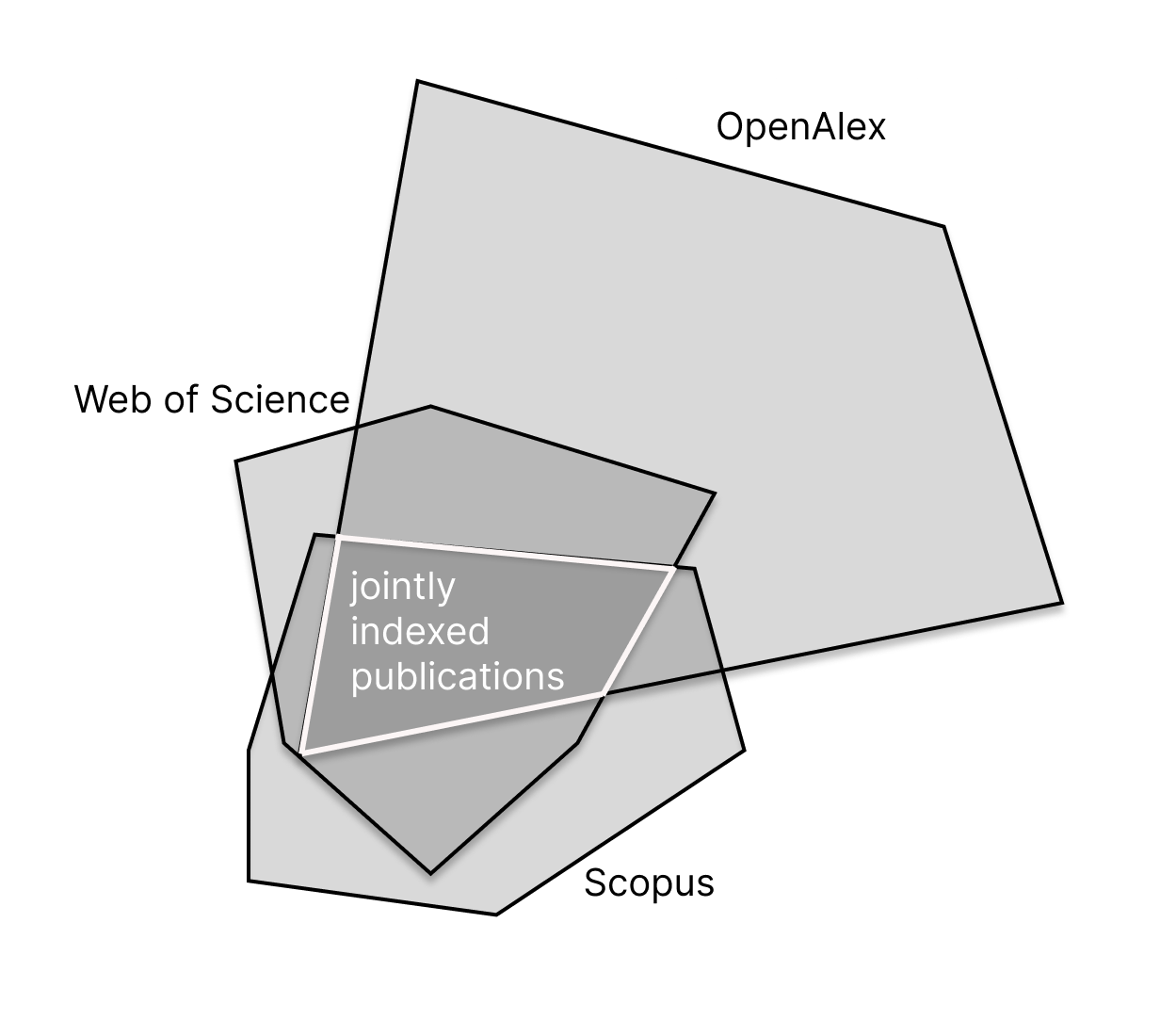}
\caption{The jointly indexed publication dataset of Web of Science, OpenAlex and Scopus, counts 15 471 336 publications. It is obtained on an exact DOI match, when publications are published between 2015 and July 2023, and where the DOI is unique. Moreover, only \textit{article} or \textit{review} in \textit{journal} are accounted.}\label{fig:1}
\end{figure}

\section{Data Collection}

The data collection process for this study entails gathering publications from three data providers: Web of Science, OpenAlex and Scopus. All the data was obtained from the German Kompetenznetzwerk Bibliometrie. The WoS and Scopus data are snapshots taken in July 2023 while the OpenAlex database is the version that was made available in December 2023. As sketched in Fig.\ref{fig:1}, we have created a \textit{jointly indexed publication} subset with records common to all three datasets based on an exact DOI match, which have been published between 2015 and July 2023, and where the DOI is unique to the record in all three databases. This allows for a comparison among OpenAlex, WoS, and Scopus controlling for the varying coverage.
The publication time window has been chosen as follows. The lower bound of the data set time is due to when the UN Sustainable Development Summit in New York approved the 2030 Agenda for Sustainable Development. Regarding the upper bound, the reason depends on our current resources' availability (i.e., the limitation due to classified publications provided by WoS, updated to July 2023). 
Overall, we include those publications classified by Web of Science as either \textit{article} or \textit{review} in \textit{journal}, and provided of an abstract (pivotal to pursue our analysis). According to the settings above, the jointly indexed publications dataset - independently of any SDG classification - of WoS, OpenAlex and Scopus, counts 15 471 336 items.

For each bibliometric database, the classified publications have been collected as follows: Clarivate directly delivered to us the WoS unique identifiers and the related SDG classifications. The OpenAlex snapshot present in our infrastructure is given of SDG labels associated with a score between $0$ and $1$; in line with their guidelines, a publication with a score above $0.4$ is considered classified with respect to the SDG in question. Regarding Scopus, Elsevier provides long lists of search queries (coded in SciVal's language) to classify publications, involving fields like: author + keywords, abstracts, journal sources, etc. We translated them into SQL and retrieved those publications by means of our infrastructure.
\begin{figure}[ht]
    \centering
    \begin{subfigure}[b]{0.32\textwidth}
        \includegraphics[width=\textwidth]{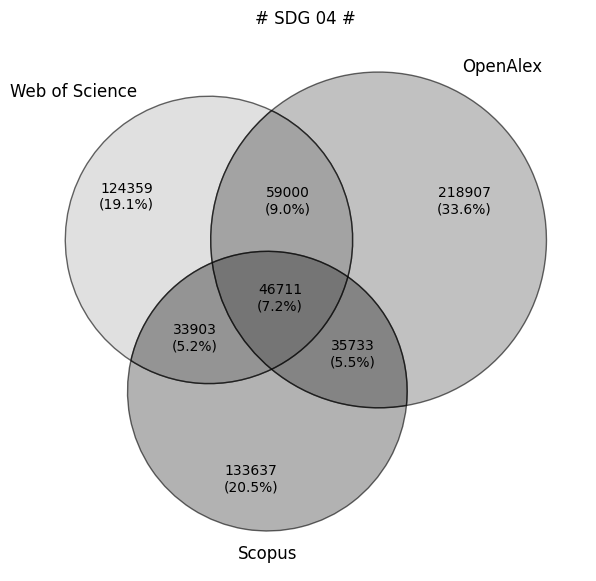}
        \label{fig:sdg4}
    \end{subfigure}
    \hfill 
    \begin{subfigure}[b]{0.32\textwidth}
        \includegraphics[width=\textwidth]{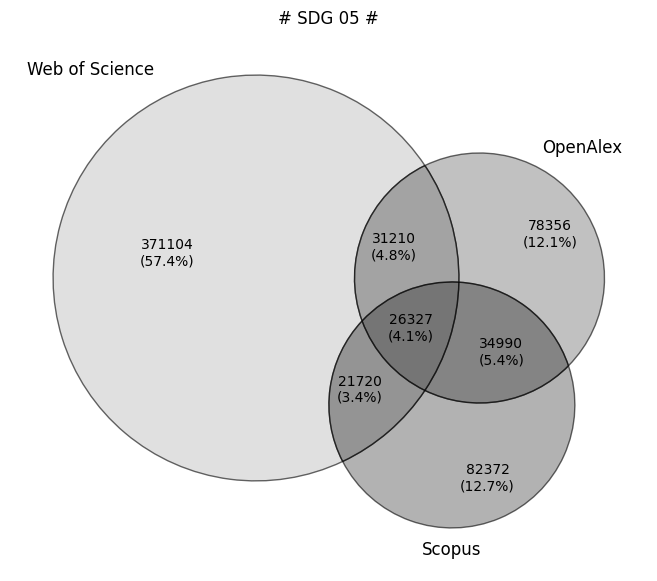}
        \label{fig:sdg5}
    \end{subfigure}
    \hfill 
    \begin{subfigure}[b]{0.32\textwidth}
        \includegraphics[width=\textwidth]{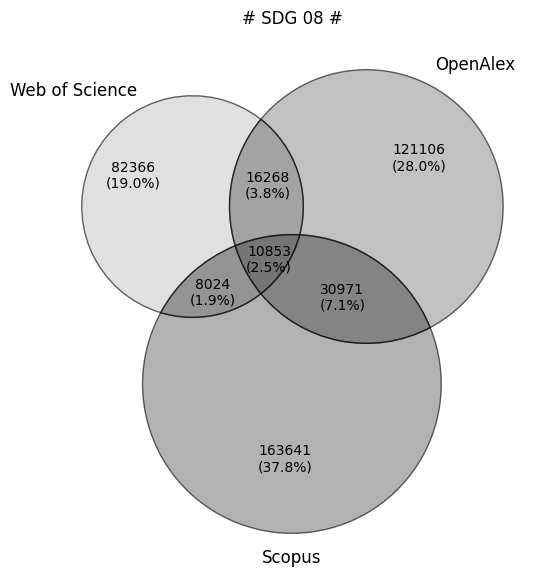}
        \label{fig:sdg8}
    \end{subfigure}
    
    \begin{subfigure}[b]{0.32\textwidth}
        \includegraphics[width=\textwidth]{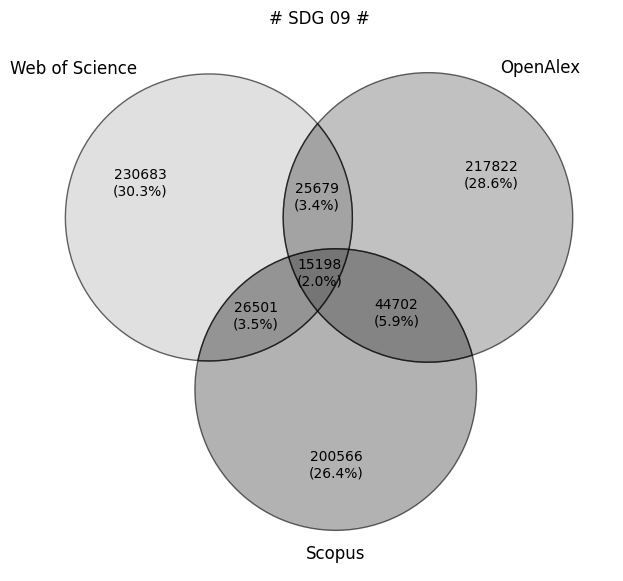}
        \label{fig:sdg9}
    \end{subfigure}
    \begin{subfigure}[b]{0.32\textwidth}
        \includegraphics[width=\textwidth]{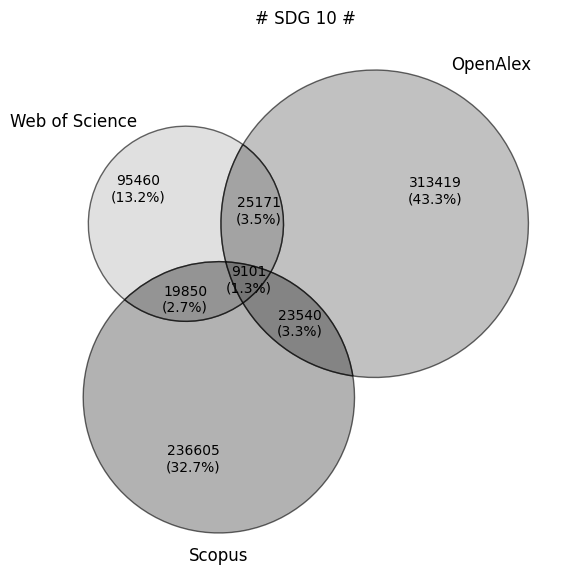}
        \label{fig:sdg10}
    \end{subfigure}
    
    \caption{Venn Diagrams of SDGs 4, 5, 8, 9, and 10, for Web of Science, OpenAlex, and Scopus.}
    \label{fig:venn_diagrams_sdgs}
\end{figure}
In Fig.\ref{fig:venn_diagrams_sdgs} we show Venn diagrams to quantify the eventual overlap of the diverse SDG classifications, reporting number of publications and percentage with respect to the total number of classified publications for each SDG. Less than $8\%$ of publications are uniformly assigned to a SDG validating former observations of highly disagreeing classifications.

\section{Methods}

The methodology employed in this research consists of two main steps: fine-tuning the DistilGPT-2 language model for each data provider and utilizing text analysis techniques to measure discrepancies among the data providers' responses to the same prompts.
DistilGPT-2 is an English speaking, faster, and lighter variant of GPT-2; the aim of its development was to provide researchers with a playground to better understand larger generative language models \citep{hugginfacesdistilgpt2}. And for that reason, it had been trained on a very limited dataset to embed the least possible prior knowledge in either content or instructions. It means that the structural data-bias of DistilGPT-2 is reduced to the minimum, letting us measure what comes from the different classifications.
For each SDG classification and each bibliometric DB, a blank copy of the DistilGPT-2 model is fine-tuned by the abstracts of those publications classified under that given SDG. This exposure allows the model to familiarize itself with the language and concepts present in the dataset, enabling it to generate responses that reflect the inherent perspectives within the data \citep{whitley2000intellectual}.

\begin{figure}
\centering
\includegraphics[width=0.75\linewidth]{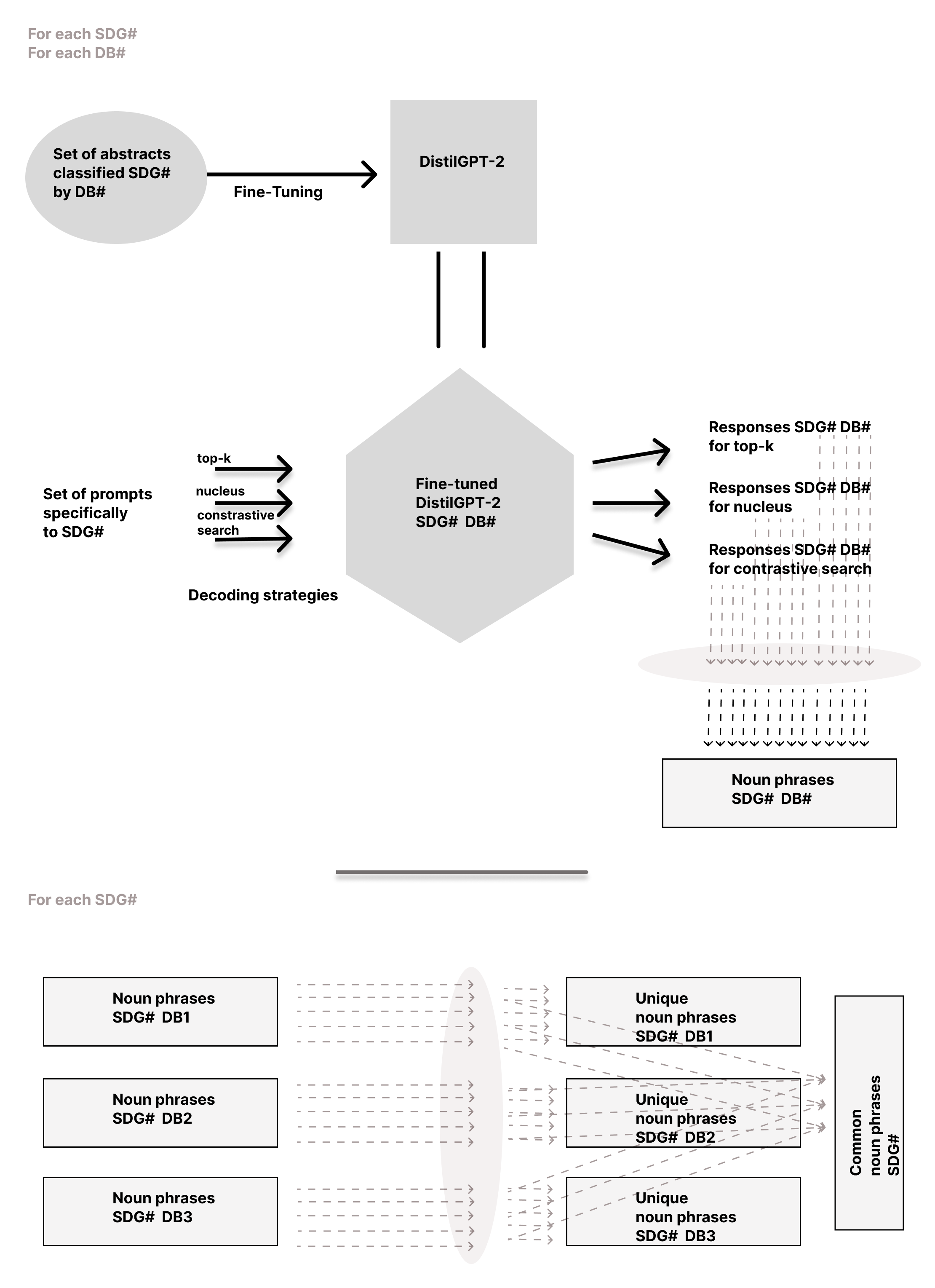}
\caption{ Schematic illustration of the research design followed in this paper. We fine-tune a blank large language model based on the architecture DistilGPT-2 to the subset of publication abstracts classified to a given SDG by a given bibliometric DB. Once obtained the fine-tuned LLM, we administrate to it a set of prompts (tailored on the SDG) through three different decoding strategies. Then, we collect into the same set the noun phrases extracted from the three response sets according to a minimum threshold. For each SDG, once obtained the latter sets for all the DBs involved, we filter out the \textit{common} words, gathering them into another set.}\label{fig:researchdesign}
\end{figure}

As outlined in Fig.\ref{fig:researchdesign}, for each SDG the same set of prompts is administered to the three fine-tuned LLM models. In particular, for each SDG, a set of prompts has been generated asking Open AI ChatGPT to produce them, by relying on the list of the official UN SDG targets. The latter methodology has been chosen because OpenAI ChatGPT might reasonably mirror an average sample of high educated individuals / experts on SDGs formulating questions. The models generate responses based on these prompts, which serve as the input for the subsequent text analysis. In order to ensuring even conditions, three decoding strategies are employed, bringing to three different responses each prompt. The latter are: \textit{top-k}, \textit{nucleus}, and \textit{contrastive search}; the first two are usually favoured by automatic evaluation while the third one by human evaluation \citep{su2022empirical,su2022contrastive}.
We analyse the responses through noun phrases analysis and topic modelling (i.e., LDA). The former resulted to be more informative and interpretable. Therefore, we extract noun phrases for each fine-tuned LLM. 
Once we obtain for each LLM a set of noun phrases alongside their occurrence, we proceed as follows:
\begin{itemize}
    \item[1] For each pair of SDG and bibliometric DB, we aggregate the noun phrases of the 3 subsets corresponding to the different decoding strategies. For each strategy, the noun phrases with a frequency higher than $10\%$ are selected. The outcome is a set of noun phrases for a given SDG classified by a given bibliometric DB.
    \item[2] For each SDG, we compare the three sets of noun phrases belonging to Web of Science, OpenAlex and Scopus. We issue a \textit{common} subset, where we collect those noun phrases common to the three bibliometric DB’s, which cannot differentiate between databases. After this filtering process, what is left is a subset of noun phrases \textit{unique} to one or any two DBs.
\end{itemize}


\begin{figure}
\centering
\includegraphics[width=1\linewidth]{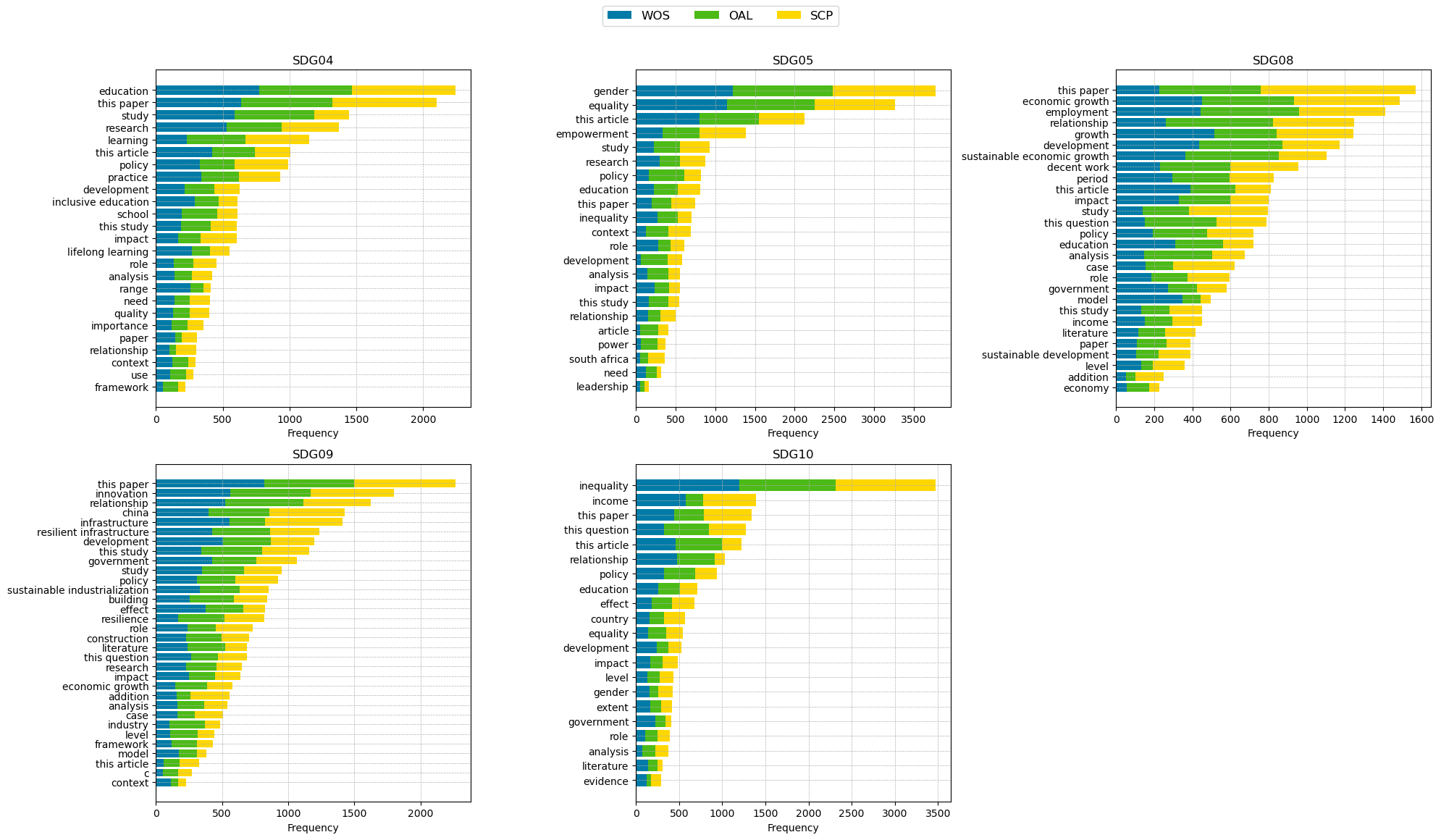}
\caption{For each SDG, the common set is the collection of those noun phrases which emerge from LLM responses in all the three databases.}
\label{fig:common}
\end{figure}

\section{Results}

\begin{figure}
\centering
\includegraphics[width=0.85\linewidth]{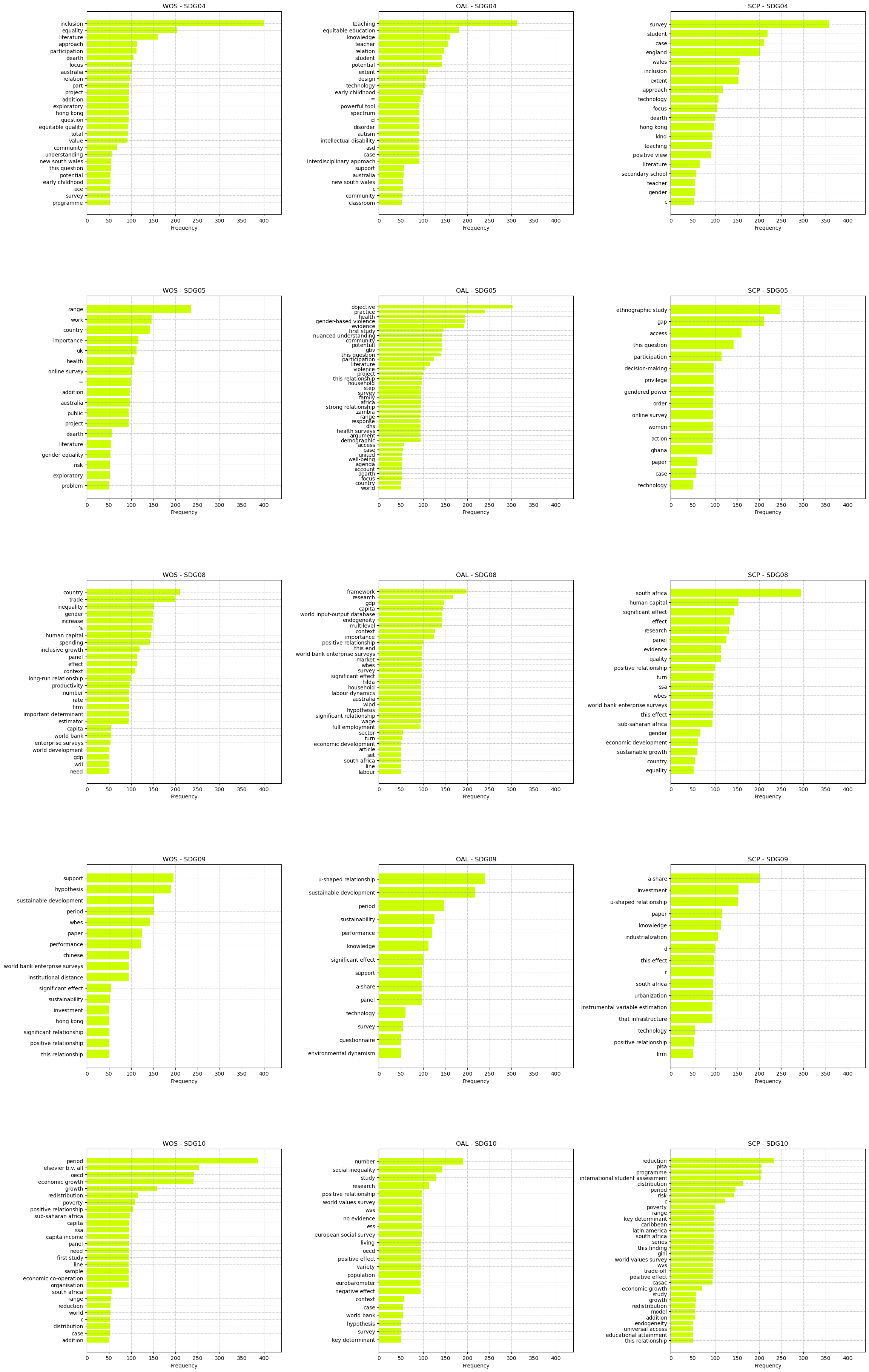}
\caption{For each SDG and bibliometric DB, frequency bar charts of the "unique" sets; i.e., noun phrases sets except those items that are in common among the three databases.}
\label{fig:unique}
\end{figure}

Overall, we fine-tune 15 LLMs to specific collections of abstracts. Each of them undergoes 500 prompts (drafted specifically to each SDG) through three decoding strategies, hence 1500 responses. 
We totally explore 1500 x 3 (DB’s) x 5 (SDGs) = 22 500 responses through noun phrases analysis. In Fig.\ref{fig:researchdesign} it is schematised the process that brings one of the 15 fine-tuned LLMs to its corresponding set of noun phrases, and then, how the outcomes from WoS, OpenAlex, and Scopus are compared for each SDG. As the prompt set counts 500 questions, and then 500 answers for each decoding strategy, then a noun phrase is considered valid in case its frequency (i.e. its occurrence among the responses) within a given strategy is greater that 50 ($10\%$). 
For each SDG, we finally obtain a set of those noun phrases found in common among the bibliometric DB’s (see Fig.\ref{fig:common}), and three sets of noun phrases belonging to each bibliometric DB, respectively assigned to any two of the three databases (see Fig.\ref{fig:unique}).
Issuing "unique" and “common” sets is particularly helpful towards the final aim of this work, i.e., assessing the data perspective of bibliometric databases in classifying SDGs.


\section{Discussion of the results}

The aim of this work is identifying data biases within the SDG classification operated by bibliometric databases through the emerging technology of large language modeling. This approach is pretty promising as of LLM capabilities, and the increasing usage in research practice and society at large lets us figuring these data biases definitely relevant to assess.
The first direct assessment of our contribution relates to identifying the publications each bibliometric database classified as a given SDG. We examine 5 SDGs. After having built a jointly indexed publication set to provide even conditions for the comparison, we have collected the publications and quantified the overlapping of the diverse classifications.

As shown in Fig.\ref{fig:venn_diagrams_sdgs}, for each SDG, the overlaps of classified publications might be extremely small, ranging from a minimum of $1.3\%$ (SDG 10) to a maximum of $7.2\%$ (SDG 4) when considering the simultaneous overlap of all three DBs. Even considering additionally the pairwise overlapping, the uncovered publications (i.e., those ones classified by only one DB) are at least ~ $73\%$ of the total in the "most agreeing classification" (SDG 4), and around $90\%$ in the most misaligned case (i.e. SDG 10). 
Such a wide difference in the publication subsets is fairly expected to produce significantly diverse outcomes among the three bibliometric DBs and highlights the performativity of the SDG classification. 

Performing fine-tuning process has contributed as first to raise our awareness of LLM usage. We have observed that different settings and hyper-parameter choices (e.g., batch size, epochs, token size, etc.) result in high sensitivity of LLM response.
Once a LLM is fine-tuned on the classification of a given SDG by a certain DB, it is then questioned through a set of prompts tailored to the corresponding SDG. 

The production of responses goes through the so-called decoding strategy. The latter determines the method by which a language model chooses the subsequent token (e.g., a word) in a sequence, based on the predicted probability for all potential tokens. Its choice has a substantial influence on the quality and diversity of the text produced.

We clearly observe how different strategies produce volatile content and text structure while maintaining high-quality responses. This remarkably enriches our awareness regarding the usage of LLM, either for generic purposes or in the research practice. This finding poses relevant ethical concerns as to this actual arbitrariness in the settings which are often thought as \textit{neutral}.

We have gathered in Fig.\ref{fig:common} frequency bar charts of the common words found each SDG explored. The most populated one is SDG 9 (33 words), while the least is SDG 5 (22 words). Despite this variation, we assess a similar shape in the distribution to some extent, somehow drawing a concave triangular matrix. 
This pattern, almost repeated each SDG, lets figuring a certain regularity deriving from the relationship between LLM and the fine-tuned data. This assures that the research design brings the three bibliometric DB. 
Conversely, in Fig.\ref{fig:unique} we show frequency bar charts of the unique sets, for each SDG and bibliometric DB. Each row corresponds to a given SDG and each column to a certain bibliometric DBs. Quantitatively speaking, SDG9 is the case we find highest agreement among the DB's. The number of words found in common is the highest and the unique word sets are the smallest. An interesting case is SDG 5, where OpenAlex identifies the double of noun phrases compared to WoS and Scopus, and the number of common words is the least.

\section{Conclusions}

The SDG classifications of WoS, OpenAlex and Scopus provide each a different perspective on what constitutes SDG. Bibliometric classifications, while striving to offer objective measures, seem to present a specific focus, which is crucial in the attribution of social relevance via SDG classifications. Depending on the applied classification scientists and institutions working in the aforementioned fields might, or might not, be able to empirically underline their impact to policy makers.

LLMs have been instrumental in unearthing and understanding these perspectives.
More important, assessing LLMs responses w.r.t. SDG classifications, lets us imagine what might lead informed decisions in the realm of policy making. 
The pre-trained DistilGPT-2 model, while smaller, offers several computational advantages; it is suitable for fine-tuning and generating scientific text. Moreover, contrary to more complex LLMs (e.g. \textit{Falcon}), it owns no knowledge about SDGs in general by default. That is an essential feature for informing about data biases after fine-tuning.

The key findings of this work might be wrap up as follows. Large Language Models are sensitive to differences in SDG classifications and provide insights into the varied perspectives inscribed into the classifications. They further show high sensitivity to model architecture, fine-tuning process and decoding strategy.

Our results clearly show how decisive is an apparently objective science-informed practice as the bibliometric classification of SDGs. A variation of classified publications and/or technical settings at various stages might widely influence the attention of society.

\section*{Funding Acknowledgments}
Data for WoS, OpenAlex, and Scopus in this study were obtained from the German Kompetenznetzwerk Bibliometrie (https://bibliometrie.info/), funded by the German Federal Ministry for Education and Research (BMBF) with grant number 16WIK2101A. The study is funded by the German Federal Ministry for Education and Research (BMBF) with grant number 01PH20006B.







\bibliography{sample}
\bibliographystyle{chicago}

\end{document}